\documentclass[twocolumn]{aastex701}
\usepackage{amsmath}
\usepackage{xcolor}

\newcommand\lat{\textit{Fermi}-LAT~}
\newcommand\source{NGC~4278~}
\graphicspath{{./}{figures/}}

\definecolor{blush}{rgb}{0.87, 0.36, 0.51}

\begin{document}

\title{VERITAS Observations Contemporaneous with the LHAASO Detection of NGC 4278}

\author{A.~Archer}\email{averyarcher@depauw.edu}\affiliation{Department of Physics and Astronomy, DePauw University, Greencastle, IN 46135-0037, USA}
\author[0000-0002-3886-3739]{P.~Bangale}\email{priyadarshini.bangale@temple.edu}\affiliation{Department of Physics, Temple University, Philadelphia, PA 19122, USA}
\author[0000-0002-9675-7328]{J.~T.~Bartkoske}\email{joshua.bartkoske@utah.edu}\affiliation{Department of Physics and Astronomy, University of Utah, Salt Lake City, UT 84112, USA}
\author[0000-0003-2098-170X]{W.~Benbow}\email{wbenbow@cfa.harvard.edu}\affiliation{Center for Astrophysics $|$ Harvard \& Smithsonian, Cambridge, MA 02138, USA}
\author[0000-0001-6391-9661]{J.~H.~Buckley}\email{buckley@wuphys.wustl.edu}\affiliation{Department of Physics, Washington University, St. Louis, MO 63130, USA}
\author[0009-0001-5719-936X]{Y.~Chen}\email{ychen@astro.ucla.edu}\affiliation{Department of Physics and Astronomy, University of California, Los Angeles, CA 90095, USA}
\author[0000-0001-5811-9678]{J.~L.~Christiansen}\email{jlchrist@calpoly.edu}\affiliation{Physics Department, California Polytechnic State University, San Luis Obispo, CA 94307, USA}
\author{A.~J.~Chromey}\email{alisha.chromey@cfa.harvard.edu}\affiliation{Center for Astrophysics $|$ Harvard \& Smithsonian, Cambridge, MA 02138, USA}
\author[0000-0003-1716-4119]{A.~Duerr}\email{a.duerr@utah.edu}\affiliation{Department of Physics and Astronomy, University of Utah, Salt Lake City, UT 84112, USA}
\author[0000-0002-1853-863X]{M.~Errando}\email{errando@wustl.edu}\affiliation{Department of Physics, Washington University, St. Louis, MO 63130, USA}
\author{M.~Escobar~Godoy}\email{mescob11@ucsc.edu}\affiliation{Santa Cruz Institute for Particle Physics and Department of Physics, University of California, Santa Cruz, CA 95064, USA}
\author{S.~Feldman}\email{sydneyfeldman@physics.ucla.edu}\affiliation{Department of Physics and Astronomy, University of California, Los Angeles, CA 90095, USA}
\author[0000-0001-6674-4238]{Q.~Feng}\email{qi.feng@utah.edu}\affiliation{Department of Physics and Astronomy, University of Utah, Salt Lake City, UT 84112, USA}
\author[0000-0002-2636-4756]{S.~Filbert}\email{u1477296@utah.edu}\affiliation{Department of Physics and Astronomy, University of Utah, Salt Lake City, UT 84112, USA}
\author[0000-0002-1067-8558]{L.~Fortson}\email{lfortson@umn.edu}\affiliation{School of Physics and Astronomy, University of Minnesota, Minneapolis, MN 55455, USA}
\author[0000-0003-1614-1273]{A.~Furniss}\email{afurniss@ucsc.edu}\affiliation{Santa Cruz Institute for Particle Physics and Department of Physics, University of California, Santa Cruz, CA 95064, USA}
\author[0000-0002-0109-4737]{W.~Hanlon}\email{william.hanlon@cfa.harvard.edu}\affiliation{Center for Astrophysics $|$ Harvard \& Smithsonian, Cambridge, MA 02138, USA}
\author[0000-0003-3878-1677]{O.~Hervet}\email{ohervet@ucsc.edu}\affiliation{Santa Cruz Institute for Particle Physics and Department of Physics, University of California, Santa Cruz, CA 95064, USA}
\author[0000-0001-6951-2299]{C.~E.~Hinrichs}\email{claire.hinrichs@cfa.harvard.edu}\affiliation{Center for Astrophysics $|$ Harvard \& Smithsonian, Cambridge, MA 02138, USA and Department of Physics and Astronomy, Dartmouth College, 6127 Wilder Laboratory, Hanover, NH 03755 USA}
\author[0000-0002-6833-0474]{J.~Holder}\email{jholder@udel.edu}\affiliation{Department of Physics and Astronomy and the Bartol Research Institute, University of Delaware, Newark, DE 19716, USA}
\author{Z.~Hughes}\email{zdhughes@wustl.edu}\affiliation{Department of Physics, Washington University, St. Louis, MO 63130, USA}
\author[0000-0002-1432-7771]{T.~B.~Humensky}\email{humensky@umd.edu}\affiliation{Department of Physics, University of Maryland, College Park, MD, USA and NASA GSFC, Greenbelt, MD 20771, USA}
\author{M.~Iskakova}\email{miskakova@wustl.edu}\affiliation{Department of Physics, Washington University, St. Louis, MO 63130, USA}
\author[0000-0002-1089-1754]{W.~Jin}\email{wjin@astro.ucla.edu}\affiliation{Department of Physics and Astronomy, University of California, Los Angeles, CA 90095, USA}
\author[0009-0008-2688-0815]{M.~N.~Johnson}\email{mjohns56@ucsc.edu}\affiliation{Santa Cruz Institute for Particle Physics and Department of Physics, University of California, Santa Cruz, CA 95064, USA}
\author{M.~Kertzman}\email{kertzman@depauw.edu}\affiliation{Department of Physics and Astronomy, DePauw University, Greencastle, IN 46135-0037, USA}
\author{M.~Kherlakian}\email{maria.kherlakian@astro.ruhr-uni-bochum.de}\affiliation{Fakult\"at f\"ur Physik \& Astronomie, Ruhr-Universit\"at Bochum, D-44780 Bochum, Germany}
\author[0000-0003-4785-0101]{D.~Kieda}\email{dave.kieda@utah.edu}\affiliation{Department of Physics and Astronomy, University of Utah, Salt Lake City, UT 84112, USA}
\author[0000-0002-4260-9186]{T.~K.~Kleiner}\email{tobias.kleiner@desy.de}\affiliation{DESY, Platanenallee 6, 15738 Zeuthen, Germany}
\author[0000-0002-4289-7106]{N.~Korzoun}\email{nkorzoun@udel.edu}\affiliation{Department of Physics and Astronomy and the Bartol Research Institute, University of Delaware, Newark, DE 19716, USA}
\author[0000-0002-5167-1221]{S.~Kumar}\email{sajkumar@udel.edu}\affiliation{Department of Physics, University of Maryland, College Park, MD, USA }
\author{S.~Kundu}\email{skundu2@crimson.ua.edu}\affiliation{Department of Physics and Astronomy, University of Alabama, Tuscaloosa, AL 35487, USA}
\author[0000-0003-4641-4201]{M.~J.~Lang}\email{mark.lang@universityofgalway.ie}\affiliation{School of Natural Sciences, University of Galway, University Road, Galway, H91 TK33, Ireland}
\author[0000-0003-3802-1619]{M.~Lundy}\email{matthew.lundy@mail.mcgill.ca}\affiliation{Physics Department, McGill University, Montreal, QC H3A 2T8, Canada}
\author[0000-0001-9868-4700]{G.~Maier}\email{gernot.maier@desy.de}\affiliation{DESY, Platanenallee 6, 15738 Zeuthen, Germany}
\author{C.~McSorley}\email{cmcsorle@ucsc.edu}\affiliation{Santa Cruz Institute for Particle Physics and Department of Physics, University of California, Santa Cruz, CA 95064, USA}
\author[0000-0002-1499-2667]
{P.~Moriarty}\email{pmoriarty0@gmail.com}\affiliation{School of Natural Sciences, University of Galway, University Road, Galway, H91 TK33, Ireland}
\author[0000-0002-3223-0754]{R.~Mukherjee}\email{rm34@columbia.edu}\affiliation{Department of Physics and Astronomy, Barnard College, Columbia University, NY 10027, USA}
\author[0000-0002-6121-3443]{W.~Ning}\email{wning@astro.ucla.edu}\affiliation{Department of Physics and Astronomy, University of California, Los Angeles, CA 90095, USA}
\author{S.~O'Brien}\email{stephan.obrien@mcgill.ca}\affiliation{Physics Department, McGill University, Montreal, QC H3A 2T8, Canada}
\author{M.~Ohishi}\email{ohishi@icrr.u-tokyo.ac.jp}\affiliation{Institute for Cosmic Ray Research, University of Tokyo, 5-1-5, Kashiwa-no-ha, Kashiwa, Chiba 277-8582, Japan}
\author[0000-0002-4837-5253]{R.~A.~Ong}\email{rene@astro.ucla.edu}\affiliation{Department of Physics and Astronomy, University of California, Los Angeles, CA 90095, USA}
\author[0000-0003-3820-0887]{A.~Pandey}\email{u6059187@utah.edu}\affiliation{Department of Physics and Astronomy, University of Utah, Salt Lake City, UT 84112, USA}
\author{C.~Poggemann}\email{connor.poggemann@utah.edu}\affiliation{Physics Department, California Polytechnic State University, San Luis Obispo, CA 94307, USA}
\author[0000-0001-7861-1707]{M.~Pohl}\email{martin.pohl@uni-potsdam.de}\affiliation{Institute of Physics and Astronomy, University of Potsdam, 14476 Potsdam-Golm, Germany and DESY, Platanenallee 6, 15738 Zeuthen, Germany}
\author[0000-0002-0529-1973]{E.~Pueschel}\email{elisa.pueschel@astro.ruhr-uni-bochum.de}\affiliation{Fakult\"at f\"ur Physik \& Astronomie, Ruhr-Universit\"at Bochum, D-44780 Bochum, Germany}
\author[0000-0002-4855-2694]{J.~Quinn}\email{john.quinn@ucd.ie}\affiliation{School of Physics, University College Dublin, Belfield, Dublin 4, Ireland}
\author[0000-0002-5104-5263]{P.~L.~Rabinowitz}\email{p.l.rabinowitz@wustl.edu}\affiliation{Department of Physics, Washington University, St. Louis, MO 63130, USA}
\author[0000-0002-5351-3323]{K.~Ragan}\email{ragan@physics.mcgill.ca}\affiliation{Physics Department, McGill University, Montreal, QC H3A 2T8, Canada}
\author{P.~T.~Reynolds}\email{josh.reynolds@mtu.ie}\affiliation{Department of Physical Sciences, Munster Technological University, Bishopstown, Cork, T12 P928, Ireland}
\author[0000-0002-7523-7366]{D.~Ribeiro}\email{ribei056@umn.edu}\affiliation{School of Physics and Astronomy, University of Minnesota, Minneapolis, MN 55455, USA}
\author{L.~Rizk}\email{leandro.rizk@mail.mcgill.ca}\affiliation{Physics Department, McGill University, Montreal, QC H3A 2T8, Canada}
\author{E.~Roache}\email{eroache@cfa.harvard.edu}\affiliation{Center for Astrophysics $|$ Harvard \& Smithsonian, Cambridge, MA 02138, USA}
\author[0000-0003-1387-8915]{I.~Sadeh}\email{iftach.sadeh@desy.de}\affiliation{DESY, Platanenallee 6, 15738 Zeuthen, Germany}
\author[0000-0002-3171-5039]{L.~Saha}\email{lab.saha@cfa.harvard.edu}\affiliation{Center for Astrophysics $|$ Harvard \& Smithsonian, Cambridge, MA 02138, USA}
\author[0009-0000-0295-8800]{H.~Salzmann}\email{hesalzma@ucsc.edu}\affiliation{Santa Cruz Institute for Particle Physics and Department of Physics, University of California, Santa Cruz, CA 95064, USA}
\author{M.~Santander}\email{jmsantander@ua.edu}\affiliation{Department of Physics and Astronomy, University of Alabama, Tuscaloosa, AL 35487, USA}
\author{G.~H.~Sembroski}\email{sembrosk@physics.purdue.edu}\affiliation{Department of Physics and Astronomy, Purdue University, West Lafayette, IN 47907, USA}
\author[0000-0002-9856-989X]{R.~Shang}\email{r.y.shang@gmail.com}\affiliation{Department of Physics and Astronomy, Barnard College, Columbia University, NY 10027, USA}
\author[0000-0003-3407-9936]{M.~Splettstoesser}\email{mspletts@ucsc.edu}\affiliation{Santa Cruz Institute for Particle Physics and Department of Physics, University of California, Santa Cruz, CA 95064, USA}
\author[0000-0002-9852-2469]{D.~Tak}\email{takdg123@gmail.com}\affiliation{SNU Astronomy Research Center, Seoul National University, Seoul 08826, Republic of Korea.}
\author{A.~K.~Talluri}\email{telid001@umn.edu}\affiliation{School of Physics and Astronomy, University of Minnesota, Minneapolis, MN 55455, USA}
\author{I.~Thoreson}\email{ithoreso@calpoly.edu}\affiliation{Physics Department, California Polytechnic State University, San Luis Obispo, CA 94307, USA}
\author{J.~V.~Tucci}\email{jtucci@iu.edu}\affiliation{Department of Physics, Indiana University Indianapolis, Indianapolis, Indiana 46202, USA}
\author[0000-0002-8090-6528]{J.~Valverde}\email{valverde@llr.in2p3.fr}\affiliation{Department of Physics, University of Maryland, Baltimore County, Baltimore MD 21250, USA and NASA GSFC, Greenbelt, MD 20771, USA}
\author[0000-0003-2740-9714]{D.~A.~Williams}\email{daw@ucsc.edu}\affiliation{Santa Cruz Institute for Particle Physics and Department of Physics, University of California, Santa Cruz, CA 95064, USA}
\author[0000-0002-2730-2733]{S.~L.~Wong}\email{samantha.wong2@mail.mcgill.ca}\affiliation{Physics Department, McGill University, Montreal, QC H3A 2T8, Canada}
\author{T.~Yoshikoshi}\email{tyoshiko@icrr.u-tokyo.ac.jp}\affiliation{Institute for Cosmic Ray Research, University of Tokyo, 5-1-5, Kashiwa-no-ha, Kashiwa, Chiba 277-8582, Japan}
\collaboration{100}{The VERITAS Collaboration}

\correspondingauthor{J.~L.~Christiansen}\email{jlchrist@calpoly.edu}\affiliation{Physics Department, California Polytechnic State University, San Luis Obispo, CA 94307, USA}
\correspondingauthor{S.~O'Brien}\email{stephan.obrien@mcgill.ca}\affiliation{Physics Department, McGill University, Montreal, QC H3A 2T8, Canada and Arthur B. McDonald Canadian Astroparticle Physics Research Institute, 64 Bader Lane, Queen's University, Kingston, ON Canada, K7L 3N6}
\correspondingauthor{M.~Pohl}\email{martin.pohl@uni-potsdam.de}\affiliation{Institute of Physics and Astronomy, University of Potsdam, 14476 Potsdam-Golm, Germany and DESY, Platanenallee 6, 15738 Zeuthen, Germany}

\begin{abstract}
Significant gamma-ray emission between 1 TeV and 20 TeV from a point source, 1LHAASO J1219+2915, consistent with the location of the LINER/LLAGN galaxy NGC 4278 was recently reported by the LHAASO collaboration.
These data were later split into active and quasi-quiet states, with most of the LHAASO significance coming from the active state (MJD 59449-59589). Subsequent analysis of \textit{Fermi}-LAT and \textit{Swift}-XRT observations have been used to explore the double-peaked broad-band emission.  
Models of the spectral energy distribution (SED) are currently unconstrained due to the lack of contemporaneous multi-wavelength data at either peak. Here we report serendipitous observations of NGC 4278 with VERITAS, made possible by the contemporaneous observations of the nearby blazars 1ES 1218+304, 1ES 1215+303, and W Comae, each of which are located within $2^\circ$ of NGC 4278. VERITAS did not detect any gamma-ray emission and a flux upper limit was calculated.  The 
flux upper limits constrain the photon spectrum of the quasi-quiet period, and together with \textit{Fermi}-LAT, indicate a peak in the SED between 100 GeV and 2 TeV.
We present an interpretation of the broadband SED that is based on acceleration of protons in the corona of the AGN, followed by p-$\gamma$ interactions and optically thin $\gamma$-ray emission.  Within this framework, the implied neutrino signal is slightly below the current sensitivity of IceCube.
\end{abstract}


\section{Introduction}
 Very-high-energy (VHE) gamma-ray emission from the nearby galaxy NGC 4278 was recently reported by the Large High Altitude Air Shower Observatory (LHAASO) in their first catalog of gamma-ray  sources (1LHAASO) \citep{LHAASO_Catalog}. The association of TeV gamma-ray emission with NGC 4278 is unexpected because GeV to TeV emitting active galactic nuclei (AGN) tend to be of the blazar or radio-galaxy class, both of which have highly relativistic jets. 
 
 NGC 4278, located at a redshift of $z = 0.002165$ \citep{redshift} is classified as a common low-ionization nuclear emission-line region (LINER) galaxy with the broad-band spectrum of a low-luminosity active galactic nucleus (LLAGN) \citep{2005ApJ...622..178G}. LINER galaxies have emission lines at optical wavelengths from neutral or ionized atoms, such as oxygen, nitrogen, and sulfur. In particular, NGC 4278 has a broad H$\alpha$ emission line indicating the existence of hot gas in the galaxy's nucleus. LINER galaxies are also identified as a low-luminosity variant of Seyfert galaxies \citep{1980A&A....87..152H}. Gamma-ray emission is only rarely associated with Seyfert galaxies; only 3 galaxies of type Seyfert (sey) are identified in the 4$^{th}$ \textit{Fermi}-LAT Catalog \citep{Fermi-Catalog1}.
 
 The source of the synchrotron emission in LLAGN is a matter of debate, attributed to either a thin accretion disk in a low-luminosity Seyfert  or a radiatively inefficient accretion flow (RIAF). Interestingly, the X-ray flux of NGC 4278 is variable.  When its X-ray flux is low, its spectral energy distribution (SED) resembles a LINER galaxy; and when it is high, it resembles a Seyfert galaxy \citep{2010A&A...517A..33Y}. 
 
Compact plasmas that emit synchrotron radiation at a sufficient rate to also support inverse-Compton scattering to GeV and TeV energies are typically found near the core of AGN within highly relativistic jets that have radio signatures too bright to be  classified as an LLAGN.  The apparent extent of the radio structure in NGC 4278 is just ${\sim}3$ pc \citep{2005ApJ...622..178G}. In addition, a low-luminosity compact radio source is identified at the core of the galaxy. These small-scale radio features, classified as compact symmetric objects (CSO), are not uncommon in LLAGN. 
\citep{general, CSO-Tremblay, CSO-Catalog}.

\citet{2005ApJ...622..178G} presented VLBI imaging of NGC\,4278 over a number of epochs, which permits measuring the apparent speed of knots and their time of ejection from the core. Pairing knots in the jet and counter jet with similar ejection time, they deduced a mildly relativistic jet speed of $\beta\simeq 0.75$, and an aspect angle $\theta\approx 3^\circ$. While perfectly consistent with the apparent speed of the knots, these values imply Doppler factors for the jet and counter jet differing by a factor $7$, and hence a substantial difference in flux that is not observed. Whatever the origin of the disagreement, the above suggests a substantial uncertainty in the jet motion and Doppler factor.  

Previous detections of GeV fluxes from NGC 4278 have been reported. The \textit{Fermi}-LAT transient catalog (1FLT) found a low-confidence detection of 1FLT J1219+2907 between March 5 and April 5, 2009 with a flux that is more than an order of magnitude higher than the \textit{Fermi}-LAT upper limits during the LHAASO observations. There is $\sim$30\% chance that this detection is a false positive and we have not been able to reproduce this detection with the Pass 8 Fermi Tools \citep{Fermi-Pass8}. 
Reports of gamma rays detected by EGRET and AGILE \citep{2024ApJS..271...10W, Dutta-radio-SED} can be attributed to source confusion in the field.  The EGRET 3EG and AGILE Grid catalogs detected gamma rays from W Comae which is \citep{Egret-EG3, Agile-Grid} well within the point-spread function of these instruments. Neither catalog identifies NGC 4278 as a source of gamma rays.
 
 To explore the SED of the LHAASO-WCDA detection, \citet{2024ApJS..271...10W} compiled contemporaneous X-ray and gamma-ray observations.  
 During the LHAASO campaign, the \textit{Swift}-XRT flux is elevated, and the peak of the flux is located at a frequency above $10^{18}$ Hz, consistent with the classification of an extreme high-peaked BL-Lacertae (EHBL) object. 
 The \textit{Fermi}-LAT observations provide only upper limits at GeV energies.  \citet{Bronzini} however, find three high-energy gamma-ray photons that correspond to a flux of $(1.9\pm1.3)\times10^{-8}$ TeV m$^{-2}$ s$^{-1}$ at 215 GeV \citep{BronziniPrivate}. 

Here we report serendipitous observations by the Very Energetic Radiation Imaging Telescope Array System (VERITAS) that overlap in time with the LHAASO observations. NGC 4278 is at small angular separation from three VHE blazars, 1ES 1218+304, 1ES 1215+303, and W Comae.   Although we do not detect the source, the excellent sensitivity of VERITAS constrains the flux above $\sim$280 GeV more strongly than the \textit{Fermi}-LAT observations. 

The VHE observations of NGC 4278 provide a new opportunity to study the physical processes that occur in LLAGN. The variability in X-rays and gamma rays may indicate short-lived injection effects \citep{Dutta-radio-SED} and transitional states. Comparison of the radio and gamma-ray luminosities suggests NGC 4278 is most similar to Fanaroff-Riley radio galaxies, though extrapolation from blazar luminosities is also consistent
\citep{LHAASO-Detection}.  
Several modeling scenarios have been presented for the SED of NGC 4278.  
In most of the models, the hard X-rays are interpreted as synchrotron emission that is reminiscent of an EHBL \citep{2024ApJS..271...10W, 2024ApJ...974..134L}.
The validity of these models for an LLAGN like NGC 4278 requires more exploration.

Here we discuss a corona scenario that naturally explains both the hard X-rays and gamma rays as well as the variability on a few-month time scale. Energetic protons in the corona interact with ambient X-rays to create pions.  The pions then decay to photons, neutrinos, and electrons. The electrons cool quickly by producing more X-rays via synchrotron emission, thus further boosting the p-$\gamma$ interactions. The transient nature of this emission is consistent with the production of the synchrotron X-rays in the region. 

This type of corona model was initially developed for the neutrino emission of the Seyfert galaxy NGC~1068 \citep{2022Sci...378..538I} for which the most energetic gamma rays are absorbed on account of a significant density of X-rays. NGC~4278 likely has a much smaller photon density, allowing for the observation of TeV gamma rays.  

We show for the first time that the coronae of LLAGN Seyfert galaxies may produce transient low-level emission of both gamma rays and neutrinos. Future more sensitive observations of this source class by LHAASO, the Southern Wide-field Gamma-ray Observatory \citep{swgo}, and the Cherenkov Telescope Array Observatory \citep{cta} will be important for multi-messenger physics.

\section{LHAASO Observations}
The first LHAASO Catalog included five extra-galactic point-sources in the 1-20 TeV energy range \citep{LHAASO_Catalog}.  Four of these are well-established TeV sources. Only 1LHAASO J1219+2915, was not yet identified as a TeV emitter \citep{TeVCat}. It is also not listed in the Fermi 4FGL catalog which spans the energy range from 100 MeV to 1 TeV \citep{Fermi-Catalog2}. The VHE emission from 1LHAASO J1219+2915 is attributed to NGC 4278.

Further analysis of the source included observations from 2021 March 5 to 2023 October 31 (MJD 59278 - 60248)  \citep{LHAASO-Detection}.  An active period of 140 days starting on 2021 August 23 and ending on 2022 January 10 was identified (MJD 59449-59589) when the flux was about 4 times larger than during the surrounding times.  The times in \citet{LHAASO-Detection} before and after the active period are referred to as quasi-quiet.

\section{VERITAS Observations}
\begin{table*}[t]
\begin{caption}
{VERITAS Observations and Results from NGC 4278(MJD 59280 - 59697) 
On-source showers are those reconstructed
within a circular region of radius 0.0707 centered on NGC 4278.  Background events are those falling in off-source regions offset from NGC 4278.
The relative area of the on- and off-source regions is $\alpha$.  The excess counts above the estimated background is Excess = On - $\alpha$Off.}\label{tab:flux_UL}
\end{caption}
\begin{center}
\begin{tabular}{lcccccccc}
\hline \hline
State & Exposure & On & Off & $\alpha$ & Excess & Significance & $E_0$ & Flux Upper Limit \\
      & [Hr] & [Counts] & [Counts] & & [Counts] & [$\sigma$] & [GeV] & [TeV$^{-1}$ m$^{-2}$ s$^{-1}$] \\ \hline 
Active     & 0.40 & 5 & 100 & 0.02778 & 2.2 & 1.2 & 620 & 1.0$\times10^{-7}$ \\
Quasi-Quiet & 6.53 & 28 & 1067 & 0.02825 & -2.1 & -0.9 & 680 & 8.3$\times10^{-9}$ \\ \hline
\end{tabular}
\end{center}
\end{table*}
 
VERITAS is an array of four 12-meter imaging atmospheric Cherenkov telescopes (IACTs) located at the Fred Lawrence Whipple Observatory in southern Arizona ($31^\circ 40\arcmin$N,  $110^\circ 57\arcmin$W) (\cite{VERITAS}).  The 3.5$^\circ$ field-of-view of VERITAS enables us to construct a mosaic of three VHE blazars, 1ES 1218+304, 1ES 1215+303, and W Comae, each of which is less than 2 degrees separated from NGC 4278.  In the analysis presented here, observations of these three sources, obtained during the LHAASO analysis time window, are used.  VERITAS collected 6.93 hours of observations between 2021 March 7 and 2022 April 28 (MJD 59280 - 59697).  Only 23.8 minutes of observations on 2022 January 3 (MJD 59582) coincide with the active period defined by the LHAASO analysis.
\begin{figure}[b]
        \centering        \includegraphics[width=1.0\linewidth]{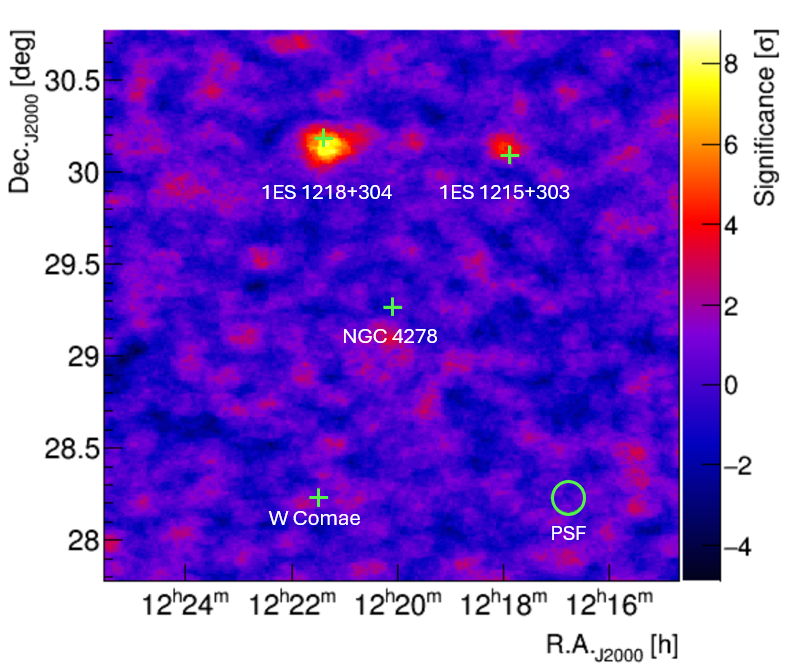}
        \caption{VERITAS sky map from observations of 1ES 1218+304, 1ES 1215+303, and W Comae contemporaneous with the LHAASO analysis.  The location of each source, including NGC 4278 is marked on the map with a \lq+' symbol and the point spread function is shown in the bottom right. There is no evidence for the detection of NGC 4278 in the VERITAS sky map. A background fluctuation with a pre-trials significance of 3.1$\sigma$ can be seen 0.15$^\circ$ below the location of NGC 4278.}
        \label{fig:skymap}
\end{figure}

Cherenkov events are reconstructed using the image template method \citep{veritas-GEO, veritas-ITM}.  The reconstruction is corrected for loss in throughput and gain due to aging of the instrument \citep{GT-factors}. We performed the analysis with cuts optimized to select gamma-ray like events from the background.  The cuts were optimized for a power law spectrum source ($F(E) = F_0 (E/E_0)^{-\Gamma}$ with $\Gamma \simeq 2.5$) using Crab Nebula data scaled to 1\% of its nominal flux. The excess counts above background are found using the reflected-region method \citep{Fomin} in an "On-source" region of radius $0.0707^\circ$ centered on NGC 4278 ($12^h 20^m 6.8^s, +29^\circ 16\arcmin 5\arcsec$). The average background is estimated from  offset, "Off-source" regions with the same radius that are normalized by $\alpha=1/N$, where $N$ is the number of regions.  The significances reported here are calculated with Equation 17 of \citet{lima} and the results were cross checked using an independent analysis package \citep{ED}.

The sky map of significance that is shown in Figure \ref{fig:skymap} was created using the ring-background method \citep{veritas-RBM}.  It includes all of the contemporaneous VERITAS data wobbled 0.5$^\circ$ from each observing target: 3.03 hours of 1ES 1218+304, 0.98 hours of 1ES 1215+303, and 3.08 hours of W Comae.  Both 1ES 1218+304 and 1ES 1215+303 appear in the maps with significances of $6.4\sigma$ and $3.7\sigma$, respectively.  No significant emission is detected at the locations of NGC 4278 ($-0.55\sigma$) and W Comae ($0.6\sigma$). W Comae was previously detected by VERITAS in both high and low states, but the low state detection required 39 hours of exposure.  \citep{WComae, WComae2}.  A background fluctuation with a pre-trials significance of 3.1$\sigma$ can be seen 0.15$^\circ$ below the location of NGC 4278.

For the SED analysis in Section 5, the data were split into two time periods, those in the quasi-quiet state, and those in the active state.  
The effective areas were calculated using the offsets between the telescope pointing and the direction of NGC 4278. This results in an average energy threshold of 240 GeV during the active period and 290 GeV during the quiet period.  The differential flux upper limits are listed in Table \ref{tab:flux_UL} at the decorrelation energy, $E_0$, where the uncertainty of the flux is smallest for different assumed values of the spectral index, $\Gamma$.

\section{\textit{Fermi}-LAT Observations}
\begin{table}[b]
\begin{center}
\begin{caption}
{\textit{Fermi}-LAT Observations: (MJD 59278 - 60248) \label{tab:flux_UL_lat}}
\end{caption}
\begin{tabular}{lcc}
\hline \hline
State & Sqrt[TS] &  Flux Upper Limit \\
      & & 100 MeV $<$ E $<$ 316 GeV \\
      & & [cm$^{-2}$ s$^{-1}$]  \\ \hline 
Active     & 2.39 & $1.8\times10^{-9}$\\
Quasi-Quiet & 3.23 & $2.1\times10^{-10}$\\ \hline
\end{tabular}
\end{center}
\end{table}
\textit{Fermi}-LAT is a space-based pair-conversion gamma-ray telescope, sensitive to gamma rays from around 20 MeV to $\gtrsim$300 GeV.
Operating in an all-sky scanning survey mode, \lat observes the entire sky approximately every 3 hours.
\lat events were reconstructed  between 59278 and 60248 MJD (636595205-720403205 mission elapsed time), in the energy range 100 MeV - 1 TeV.
Events within 15$^\circ$ of \source were considered, with a zenith cut of 90$^\circ$ applied to remove contamination due to gamma rays from the Earth's limb.
The remaining events were analyzed using the Science Tools package (\textit{fermitools} v2.2.0), using the \textit{fermipy} \citep[v1.3.0][]{fermipy} analysis suite.
``\textit{Source}'' class events (evclass=12) converting in both the front and the back of the instrument (evtype=3) were analyzed.
``Pass 8'' \citep[\textit{P8R3\_SOURCE\_V3}][]{Fermi-Pass8} instrument response functions (IRFs) were used, with the ``gll\_iem\_v07.fits'' and ``iso\_P8R3\_SOURCE\_V3\_v1.txt,'' galactic and isotropic diffuse models, respectively.

\par A binned-likelihood analysis was applied, with different event types (4, 8, 16, 32) handled separately, with their respective IRFs, using a summed-likelihood analysis.
A model was constructed including all known sources included in the 4th Fermi Point Source Catalog, data release 2 \citep[4FGL][]{Fermi-Catalog2, 4FGL-dr2}. 
Spectral shape and normalization parameters were left free to vary during the fitting process for sources which were within 5$^\circ$ of NGC 4278, or which had a test statistic (TS) $\ge$10.
In addition, the galactic and isotropic diffuse components were allowed to vary. 
The source was modeled using a point-source spatial model centered on the location of NGC 4278 and a power-law spectrum, $F=F_0 (E/E_0)^{-\Gamma}$, with a scale energy, $E_0$ of 1 GeV.
Results for the active and quasi-quiet periods are summarized in Table \ref{tab:flux_UL_lat}. 
In the absence of a detection, upper limits (95\% confidence level) on the fluxes are reported in Column 3 of Table \ref{tab:flux_UL_lat}.

\section{Spectral Energy Distribution}

\begin{figure*}[t]
    \centering   
    \includegraphics[width=1.0\linewidth] {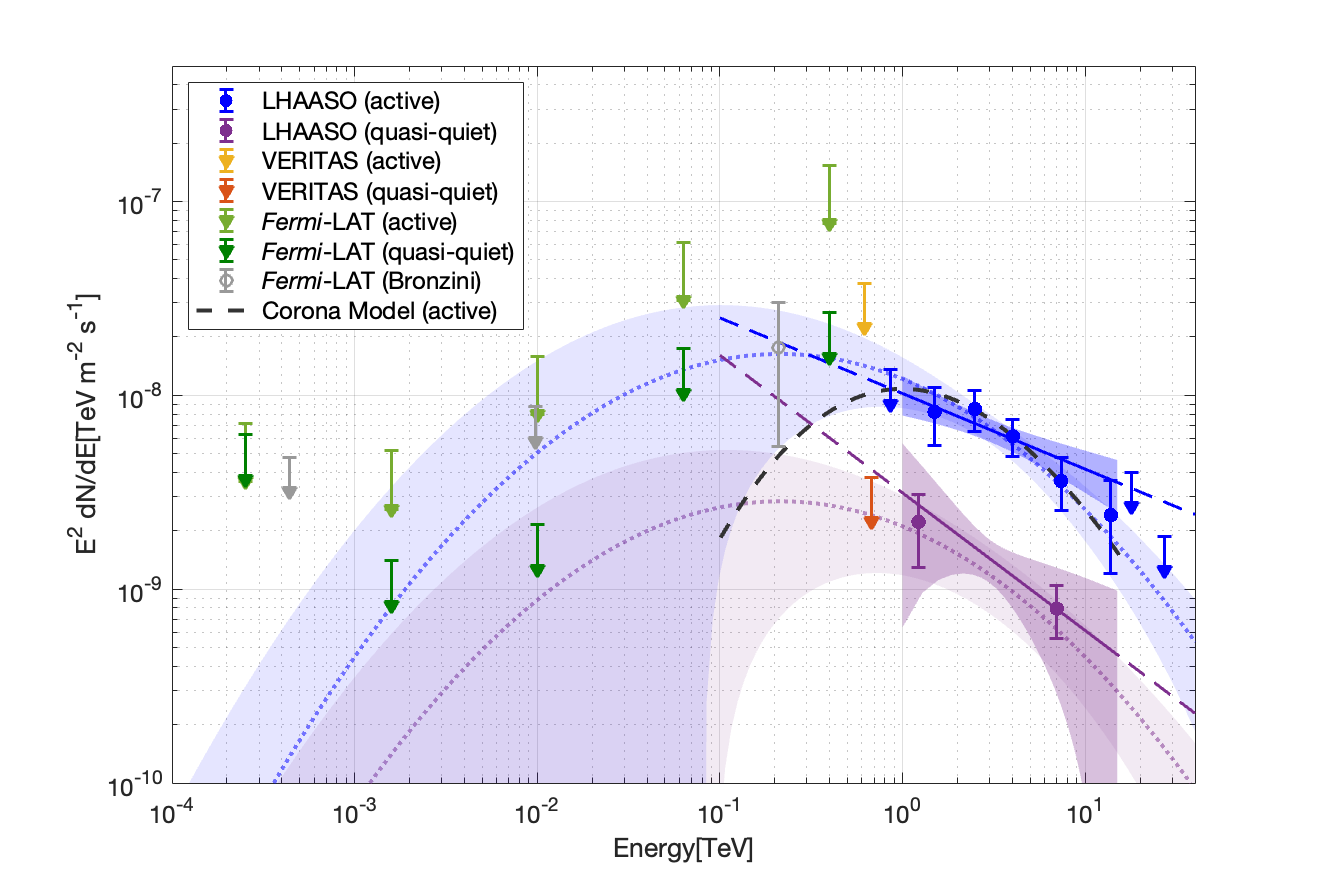}
    \caption{NGC 4278 SED during the active and quasi-quiet periods: LHAASO-WCDA fluxes and upper limits from LHAASO-KM2A (blue \& purple), VERITAS (yellow \& orange), and \textit{Fermi}-LAT (green \& light green). The \citet{Bronzini} \textit{Fermi}-LAT (gray) is from data spanning the 1LHAASO Catalog dates. The power-law fit to the LHAASO data is from \citet{LHAASO-Detection}. A log-parabolic model was fit to the combined dataset for the active period (blue dotted line) to explore the curvature of the spectrum. During the quasi-quiet period (purple dotted line), the two LHAASO-WCDA flux points do not constrain the curvature, so only the normalization was fit assuming the same spectral shape as the active period. The example corona model presented in section 6.1 is also included.}
    \label{fig:sed}
\end{figure*}

The LHAASO observations are considered in the context of the broader multiwavelength landscape of the VERITAS and \textit{Fermi}-LAT upper limits in Figure \ref{fig:sed}. 
Here we see that, during the quasi-quiet state, the VERITAS observations are mildly inconsistent with a power-law extrapolation of the LHAASO data.  Under the assumption that the VERITAS observations probe the average flux during this state, the power-law index cannot exceed 3.13 at 95\% confidence, which is lower than the maximum value of 3.40 reported by LHAASO \citep{LHAASO-Detection}. The best fit value of the index including the VERITAS upper limit is $2.57 \pm 0.23$ at 68\% confidence.  This constrains models of the VHE emission more strongly than the \textit{Fermi}-LAT upper limits at these energies. The single VERITAS run during the active period is not enough to constrain the LHAASO power law. It is worth noting that most of the VERITAS observations were taken during the quasi-quiet period, and most of the LHAASO flux was recorded during the active period, which invalidates a comparison of a combined VERITAS upper limit with the average LHAASO spectrum. 

The LHAASO fluxes and upper limits from \textit{Fermi}-LAT during the active period suggest that the spectrum has a peak below or possibly just above the energy threshold of the LHAASO-WCDA detector as shown in Figure \ref{fig:sed}. The curvature of inverse-Compton emission in blazars is typically modeled with a log-parabolic function, $F(E) = F_0(E/E_0)^{-\Gamma-b\ln(E/E_0))}$, where the amount of curvature is determined by the parameter $b$.
For these data, we set the scale energy, $E_0 $, to 3.0 TeV, the approximate energy where the LHAASO-WCDA sensitivity is maximal. 
The parameters of the log-parabolic function were determined using a least-squares fit to the SED.  The VERITAS upper limit was included by assuming that it represents a flux of zero, with an uncertainty that is half of the flux upper limit.  The \textit{Fermi}-LAT analysis provides spectral fluxes even when they are of low significance, and these were included in the fit.
The fit results in a flux normalization of $F_0 = (7.6 \pm 1.3)\times 10^{-10}$ photons TeV$^{-1}$ cm$^{-2}$ s$^{-1}$ with a spectral index of $\Gamma =2.66 \pm 0.21$ and a curvature term, $b = 0.125 \pm 0.048$ with parameter uncertainties reported in a 95\% confidence interval. 
Reporting the uncertainties at 95\% confidence allows us to explore the possible shapes of the VHE peak that are consistent with the LHAASO spectrum. 
The covariance matrix shows that the curvature parameter is strongly correlated with the spectral index. After carefully considering the uncertainties in the fitted parameters, and the systematics of our technique, we conclude that the peak in the flux may be located anywhere between 100 GeV and 2 TeV.

During the quasi-quiet period, there are only two LHAASO flux points, which is not enough to provide reasonable constraints on the curvature of the spectrum.  Instead, we fix the index and curvature to the values found previously for the active period and only allow the flux normalization to vary.  The flux normalization during the quasi-quiet period is $F_0 = (1.32 \pm 0.49)\times 10^{-10}$ photons TeV$^{-1}$ cm$^{-2}$ s$^{-1}$.  This flux is 5.7 times smaller than the flux during the active period.  The butterfly shown includes the shape-preserving uncertainty on the index and curvature from the previous fit to the active period. 

For reference, the flux and upper limits reported by \citet{Bronzini} are also shown in Figure \ref{fig:sed}.  Their \textit{Fermi}-LAT analysis was derived from data spanning the 1LHAASO Catalog dates, 2021 March to 2022 October. The flux at 215 GeV is inferred from the weakly significant detection of three photons, one detected during the active period and two during the quasi-quiet period. Because it is unclear how this flux point is related to the active and quasi-quiet time periods, we did not include it in our fits.  We note however, that it lies below the upper limits in both time periods as expected, and that the uncertainty on the flux is sufficiently large to be consistent with the fits presented here.

\citet{2024ApJ...974..134L} presented a fit to the SED of a synchrotron-self-Compton (SSC) model. For a Doppler factor $D=2.7$, following \citet{2005ApJ...622..178G}, the variable optical emission \citep{2010A&A...517A..33Y} could not be reproduced, suggesting larger Doppler factors (a model for $D=10$ was shown). \citet{2024ApJS..271...10W} find reasonably good SSC fits to the SED for small and large ($30^\circ$) viewing angle of the jet, if attributing the optical emission to the host galaxy. The Doppler factor for the large aspect angle would be small, $D\simeq 1.8$. For the well-aligned jet it is $D\approx 10$. The fact that the X-ray/TeV spectrum can be reasonably well reproduced for a wide range of Doppler factors suggests that even unboosted SSC emission could provide an acceptable model. This would include particle acceleration and radiation production in a corona by directly tapping the energy in turbulence \citep[e.g.][]{2021PhRvD.104f3020L,2023JPlPh..89e1701L}. 
 
Recent models of stochastic particle acceleration in the coronae of AGN suggest efficient acceleration of protons and that neutrinos might be produced through p-$\gamma$ interactions \citep{2025A&A...697A.124L}. In the case of NGC\,1068, TeV-scale neutrino emission has indeed been observed \citep{2022Sci...378..538I} and the theoretical implications have been extensively discussed \citep[e.g.][]{2024ApJ...961L..14F,2024PASJ...76..996I}. In the following section we explore a similar scenario for NGC~4278. An example corona model is included in Fig. \ref{fig:sed}.

\section{Interpretation}
\subsection{Hadronic interactions}
In this section we analytically explore the viability of a hadronic origin of the $\gamma$-ray emission observed from NGC\,4278. \citet{2024ApJS..271...10W} and \citet{2025arXiv250702326S} have already discussed a proton-proton (pp) scenario involving inelastic collisions of fast and slow protons. We focus on photohadronic (p-$\gamma$) models, in which VHE protons interact with photons to produce pions that eventually decay into neutrinos, gamma rays, and electron/positron pairs that produce synchrotron and inverse-Compton emission. The luminosity of leptonic emission, which in the steady state is equal to the injection power of the pairs, is about the same as that of neutrinos and that of the gamma rays that result from the decay of neutral pions. Since each charged pion provides a muon neutrino, a muon antineutrino, and one type of electron neutrino, and since there are six possible states of neutrinos, complete flavor mixing reduces the muon-neutrino flux  by a factor of two during propagation to the observer.

For the corona model, no Doppler boosting is assumed\footnote{While this paper was under review, two other publications discussing leptohadronic models appeared on the arXiv, both within an outflow scenario \citep{yuan2026tevgammarayslowluminosityactive,chen2026physicaloriginveryhighenergygamma}}. The observed peak in the gamma-ray emission at a few hundred GeV can then be used to infer a pion energy of about $1$~TeV. Assuming interaction through the $\Delta$ resonance, the proton energy would be about $E_\mathrm{p}\approx 20$~TeV \citep{1995PhR...258..173G,2010ApJ...721..630H}.
The target photons would need to have an energy of 
\begin{equation}
\epsilon_\mathrm{target} \approx (200\,\mathrm{MeV}) \frac{m_\mathrm{p}c^2}{E_\mathrm{p}}
\approx 10\,\mathrm{keV}\ ,
\label{eq1}
\end{equation}
very close to the peak of the synchrotron emission in the SSC models. The rapid decline of the photon density above $10$~keV imposes a very hard spectrum of hadronic emission below $100$~GeV, both for neutrinos and for photons. In the following we only consider single-pion production with hard X-rays. This implies that the proton spectrum falls off steeply above $20$~TeV. To be noted is that the acceleration rate of protons, and hence their maximum energy, does not directly correlate with the spatial diffusion coefficient that is relevant for the escape rate. There is an unknown dependence on the mean energy gain per scattering mean free path that decouples the two, and so setting the maximum energy does not preclude the choice of escape rate. For example, \citet{Comisso_2019} find the acceleration time scale fairly independent of energy, whereas the escape time in turbulence appears to scale like a power law, $t_\mathrm{esc}\propto E^{-\delta}$, with $0.3\lesssim \delta \lesssim 0.5$ \citep{2023JPlPh..89e1701L,2023MNRAS.525.4985K}.
 
For simple one-zone radiation modeling we assume a spherical, homogeneous emission region with radius $R$. The distance to NGC\,4278 is somewhat uncertain. The nominal redshift, $z\simeq 0.002$, suggests a distance below $10$~Mpc, whereas the redshift corrected to the reference frame defined by the CMB indicates values around $13$~Mpc, as do with considerable scatter various redshift-independent estimates\footnote{\url{https://ned.ipac.caltech.edu}}. Using the latter value for the luminosity distance and the X-ray spectrum as shown in \citet{2024ApJS..271...10W} we estimate the differential X-ray luminosity, $L_\epsilon$, at the time of elevated activity as
\begin{equation}
\epsilon L_\epsilon\simeq 6\cdot 10^{40}\ \mathrm{erg\,s^{-1}} 
\label{eq2}
\end{equation}
yielding the differential photon density, $n_\epsilon$, in the source at the level
\begin{equation}
\epsilon n_\epsilon =\frac{3}{4\pi R^2 c} L_\epsilon\simeq (5\cdot 10^{7})\,R_{15}^{-2}\ \mathrm{cm^{-3}} ,
\label{eq3}
\end{equation}
where we assume optically thin conditions, $R_{15}$ denotes the radius of the emission zone in units of $10^{15}$~cm, and $c$ stands for the speed of light; we take the mean photon energy as $\epsilon =6$~keV.

The X-ray photons do not provide a significant pair-production opacity for $\gamma$ rays, unless the emission zone is far smaller than $R_{15}=0.01$. Optical photons would be far more abundant, if produced in the same emission zone, and would impose significant absorption of TeV photons. As the $\gamma$-ray emission peaks at $300$~GeV, where the absorption should be most relevant, we conclude that most of the optical emission originates outside of the emission zone and possibly in a large volume. Support for that proposition comes from the observed spectrum peaking in the optical band rather than the UV band, that is typically associated with an accretion disk, and that the optical component can be well explained by the host galaxy \citep{2024ApJS..271...10W}.

We calculated an example $\gamma$-ray spectrum and added that to Figure~\ref{fig:sed}. We used the simplified model of \citet{2010ApJ...721..630H}, in particular their eqs.~7, 27, and 30, assuming interactions through the $\Delta$ resonance only.
The X-ray spectrum is assumed to be typical of corona spectra and consistent with that observed during the LHAASO high state,  
\begin{equation}
N_X(E_X)\propto E_X^{-1.8}\,\exp\left(-\frac{E_X}{150\,\mathrm{keV}}\right),
\label{eq3a}
\end{equation}
whereas the proton spectrum is taken to be
\begin{equation}
N_p(E_p)\propto \frac{E_p^{-2.2}}{\left(1+\frac{E_p}{15\,\mathrm{TeV}}\right)^2}.
\label{eq3b}
\end{equation}
We emphasize that both the $\gamma$-ray spectrum and the input X-ray and particle spectra used to compute it really are examples and not a model prediction. The particle spectrum in particular may be more complex than this \citep{2024ApJ...974...75F}.

The absence of absorption implies that the expected neutrino spectrum would be approximately the same as that observed in $\gamma$-rays. 
Based on the gamma-ray spectrum measured during the active state of NGC 4278, we assume the neutrino spectrum follows a power law with an index $2$ and a normalization of $2\times10^{-12}\; \mathrm{TeV^{-1}\,cm^{-2}\,s^{-1}}$ at 1 TeV. Using the effective area of IceCube for gamma-ray follow-up Bronze alerts \citep{2019ICRC...36.1021B}, we compute the expected count spectrum across the IceCube energy range within the 140-day active period. The sum of the expected neutrino counts amounts to $\sim$0.2 events seen by IceCube in this period. Similarly, assuming the quasi-quiet state (normalization of $3\times10^{-13}\; \mathrm{TeV^{-1}\,cm^{-2}\,s^{-1}}$ at 1 TeV) lasted for 10 years, we estimate that roughly 0.7 events are expected in the IceCube detector over the 10 years.

The typical energy of secondary electrons and positrons coincides with the break energy in the assumed electron spectra in SSC fits \citep[cf. Table 2 of][]{2024ApJ...974..134L}, suggesting that the SSC models might be a superposition of truly leptonic emission and that of secondary electrons.

The cross section of $p\gamma$ interaction is $\sigma_{p\gamma} \approx 5\times 10^{-28}\ \mathrm{cm^2}$. With the inelasticity $K=0.2$ \citep{2000CoPhC.124..290M} we find for the energy-loss time of the radiating protons,
\begin{equation}
t_{p\gamma}\approx\frac{1}{K\,c\,\epsilon n_{\epsilon}\,\sigma_{p\gamma}}
\approx R_{15}^2\, \left(6\times 10^{9}\ \mathrm{s}\right) .
\label{eq4}
\end{equation}
Radiative-MHD simulations suggest that the corona should be supported by magnetic pressure \citep{2014ApJ...784..169J}. The time available for proton acceleration and their p-$\gamma$ interaction should not not be limited by the infall time of the accretion flow, but rather by diffusive escape from the corona.
Using the escape time, $t_\mathrm{esc}$, the radiative efficiency of the protons is given by
\begin{equation}
\eta=\frac{t_\mathrm{esc}}{t_{p\gamma}+t_\mathrm{esc}} .
\label{eq5}
\end{equation}
Assuming Bohm diffusion in a fully turbulent magnetic field of strength $B\approx 4$~G (cf. subsection~\ref{sec-se}), we find an upper limit to the escape time,
\begin{equation}
t_\mathrm{esc}\lesssim R_{15}^2\,(3\times 10^{9}\ \mathrm{s}) ,
\label{eq6}
\end{equation}
and a radiative efficiency of up to a few tens of percent. If the diffusion coefficient were a hundred times the Bohm limit, we would find for the radiative efficiency $\eta\approx 0.01$ and for the escape time about a year, commensurate with the duration of the $\gamma$-ray high state. The same radiative efficiency, $\eta\approx 0.01$, would result with Bohm diffusion and a magnetic-field amplitude as preferred by the SSC models, $B\approx 30$~mG. In both cases the implied minimal proton injection power would be $10^{43}$~erg/s. Unlike other AGN associated with multi-TeV neutrino emission \citep[e.g.][]{2019NatAs...3...88G,2023ApJ...954...70A}, there is no energetics problem for a hadronic corona model of NGC~4278.

The empirically determined maximum proton energy of $20$ ~TeV should represent a balance between the acceleration rate and either the available time during the high state or a loss process. Eq.~\ref{eq4} shows that radiative losses are very slow, and so the relevant loss process must be escape. The acceleration rate likely decreases with energy, and the escape rate increases, both on account of the energy dependence of Larmor deflection, and so we implicitely require that acceleration is the fastest process at energies below about $20$~TeV, and escape is the fastest above this energy.

\subsection{Secondary electrons}\label{sec-se}
The corona derives the energy needed for particle acceleration from turbulence generation and reconnection in the accretion flow and its interface to the emission region. The in-flux of magnetic energy should be at least as large as the observed nonthermal luminosity,
\begin{equation}
\epsilon L_\epsilon \lesssim \frac{B^2}{8\pi} c\,\pi R^2\quad
\Rightarrow\quad B \gtrsim (4\ \mathrm{G})\,R_{15}^{-1} .
\label{eq7}
\end{equation}
In p-$\gamma$ interactions, secondary electrons typically have about the same energy and source power as do the $\gamma$-rays and the neutrinos. Peak production is therefore expected at $E_\mathrm{e}\approx 500$~GeV, below which the spectrum will be hard, $N_\mathrm{e}\propto E_\mathrm{e}^{-2}$, on account of synchrotron cooling. Given a magnetic-field strength as in eq.~\ref{eq7}, the synchrotron spectrum would be hard up to 
\begin{equation}
E_\mathrm{sy, se} \gtrsim (20\ \mathrm{keV})\,R_{15}^{-1} . 
\label{eq8}
\end{equation}
Synchrotron cooling likely is the dominant cooling mechanism for electrons, as inverse-Compton scattering is Klein-Nishina suppressed. Hence the synchrotron flux at hard X-rays should be similar to the $\gamma$-ray flux at $500$~GeV, as is observed, suggesting that the p-$\gamma$ interactions enhance the abundance of their own target photons through synchrotron radiation of secondary electrons. As the synchrotron spectrum would be hard in the Swift-XRT band ($\Gamma\approx 1.5$, $N(E_X)\propto E_X^{-\Gamma}$), the Comptonization component must be softer than the total spectrum and hence is reasonably well in line with the X-ray spectra observed from unobscured type-1 AGN \citep[$1.6\lesssim\Gamma\lesssim 2.4$,][]{2012MNRAS.420.1825J}.

Electron-positron pairs are also produced through the Bethe-Heitler process ($p+\gamma\rightarrow e^+ + e^-$). Their energy is about a factor of $50$ smaller than that of the electrons from muon decay, and correspondingly their synchrotron frequency is a factor $2500$ lower, placing it in the optical band or below (cf. eq.~\ref{eq8}), where the host galaxy is bright. The cross section, multiplied with the inelasticity, is about two orders of magnitude smaller than that of $p$-$\gamma$ interactions, and there is another factor $10$ reduction in the energy transfer to the pairs for the same center-of-momentum energy of the interaction \citep{Gao_2017}. Therefore, for any energetically reasonable spectral index of protons below the maximum energy around $20$~TeV ($s=2.0-2.5$, $N_p\propto E^{-s}$), there is not enough source power in the Bethe-Heitler process to sustain a significant contribution to the SED. 
\subsection{Primary electrons}
It is very likely that primary electrons will be accelerated by the same process that also energizes protons. The maximum energy, $E_\mathrm{e,max}$, will be smaller than that of the protons, $E_\mathrm{p,max}$, on account of rapid energy losses, mostly by synchrotron radiation. \citet{2025PhRvE.112a5205L} finds that the local acceleration time can be similar to the eddy turn-over time of the turbulence. Particle acceleration is therefore an efficient damping mechanism for turbulence. The average acceleration time must allow the turbulence to fill the emission zone. Taking $10\%$ of the light-crossing time for the mean acceleration time and equating that with the synchrotron loss time yields $E_\mathrm{e,max}\lesssim 5$~GeV, and the peak energy of their synchrotron radiation would be around $2$~eV. The observed emission in the optical and NIR is very bright and can easily outshine the synchrotron radiation from primary electrons.

\section{Conclusions}
The LHAASO detection of TeV-scale $\gamma$ rays from the LLAGN \source has opened a new window on intermittent or emergent AGN activity within galaxies. We present serendipitous VERITAS observations taken during the active and the quiet states of the source. The contemporaneous \textit{Fermi}-LAT and VERITAS upper limits constrain the VHE emission of this unique source, and we infer a spectral peak (in $\nu F_\nu$) at a few hundred GeV which constitutes the third spectral component besides the prominent optical emission and the unusually hard X-ray emission. 

As SSC models and pp-scenarios have already been discussed in the literature, we explore an interpretation involving proton acceleration in the corona of the accretion flow, followed by p-$\gamma$ interactions. 

Our findings can be summarized as follows:

\begin{itemize}

\item The Doppler factors of the jet components observed with VLBI are very uncertain, and SSC modeling allows reasonable fits for Doppler factors near unity with unbeamed emission, suggesting the viability of corona models.

\item For a reasonable size of the emission zone, $R=10^{15}$~cm, equivalent to $12\,R_G$ for a black-hole mass of $M_\mathrm{BH}\simeq 3\cdot 10^8\, M_\odot$\citep{2003MNRAS.340..793W}, the pair-production opacity in the TeV band is likely well below unity, implying that the spectral peak at a few hundred GeV must be intrinsic.

\item A hadronic corona model for the VHE emission appears feasible, can naturally produce a spectral peak in the TeV band, and does not violate any fundamental requirement that we tested. In the optically thin case, hadronic scenarios predict similar spectra for neutrinos and $\gamma$ rays. At $1$~TeV the neutrino flux would be about a factor of $30$ below that observed from NGC~1068, but the spectrum would be harder by about $0.7$ in index, indicating parity in flux at about $100$~TeV, if the power-law spectra extrapolate that far. As this neutrino flux is reached only during the high state of NGC~4278, we expect the signal to be slightly below the sensitivity of IceCube, roughly at 0.2 neutrino events over the 140-day active period.

\item
Synchrotron emission from secondary electrons would peak in hard X rays with a flux similar to that observed \citep{2024ApJS..271...10W, 2024ApJ...974..134L}. This offers an explanation for the hard X-ray spectrum that is reminiscent of an EHBL and not a LLAGN. It would also suggest that the p-$\gamma$ interactions amplify themselves by boosting the abundance of the required target photons, which must be in the hard X-ray range for the production of TeV-scale $\gamma$ rays, neutrinos, and secondary electrons.

\item The acceleration of primary electrons is limited by rapid synchrotron cooling. We estimate their maximum energy to be a few GeV, leading to a synchrotron peak in the optical where other emission processes may dominate the radiative output of the source.

\end{itemize}

\section*{Acknowledgments}
This research is supported by grants from the U.S. Department of Energy Office of Science, the U.S. National Science Foundation and the Smithsonian Institution, by NSERC in Canada, and by the Helmholtz Association in Germany. This research used resources provided by the Open Science Grid, which is supported by the National Science Foundation and the U.S. Department of Energy's Office of Science, and resources of the National Energy Research Scientific Computing Center (NERSC), a U.S. Department of Energy Office of Science User Facility operated under Contract No. DE-AC02-05CH11231. We acknowledge the excellent work of the technical support staff at the Fred Lawrence Whipple Observatory and at the collaborating institutions in the construction and operation of VERITAS.  J. Christiansen acknowledges the generous support of the Marrujo Foundation.

\bibliography{references.bib}{}
\bibliographystyle{aasjournalv7}

\end{document}